# SPATIAL AND TEMPORAL SPECTRA OF NOISE DRIVEN STRIPE PATTERNS

K.Staliunas

Physikalisch Technische Bundesanstalt, 38116 Braunschweig, Germany

tel.: +49-531-5924482, Fax: +49-531-5924423, E-mail: Kestutis.Staliunas@PTB.DE

**Abstract**

Spatial and temporal noise power spectra of stripe patterns are investigated, using as a model a Swift-Hohenberg equation with a stochastic term. In particular, the analytical and numerical investigations show: 1) the temporal noise spectra are of $1/f^a$ form, where $a = 1+(3-D)/4$ with $D$ the spatial dimension of the system; 2) that the stochastic fluctuations of the stripe position are sub-diffusive.

Stripe- (or roll-) patterns appear in a variety of spatially extended systems in nature, like ripples in sand, or markings of the skins of the animals, and also in variety of physics laboratory systems, like Rayleigh-Benard convection [1], Taylor-Couette flows [2], or wide aperture nonlinear optical resonators (degenerate optical parametric oscillators [3], four wave mixing [4]). Several microscopic mechanisms for (roll) pattern formation are already understood, such as the Turing mechanism in chemical or biological systems [5], and the "off-resonance excitation" in nonlinear optical systems [6]. For the large variety of pattern forming systems, despite of their different microscopic pattern formation mechanisms, a universal description of stripe patterns is possible. Universal features of stripe pattern dynamics (e.g. the "zig-zag" or Eckhaus instabilities) are well known, as investigated on universal model equations: the Swift-Hohenberg equation as an order parameter equation for stripes in spatially isotropic system [7], or the Newell-Whitehead-Segel equation as an amplitude equation for perturbations of stripe patterns [8].

The above referenced investigations deal with dynamics of stripes in the absence of noise. It is, however, well known that noise, as present in every system, can bring about new features in the behavior of stripe patterns (and of patterns in general). First, noise can modify (shift) the very threshold of stripe formation [9]. Second, noise can lead to precursors of stripes below the pattern formation threshold [10]. While noiseless pattern forming system below the pattern formation threshold show no patterns at all, since all perturbations decay, one observes in the presence of noise a particularly spatially filtered noise, which e.g. in nonlinear optics has been named "quantum patterns", when the noise is of quantum origin [11]. Finally, above the pattern formation threshold noise can result in defects (dislocations, disclinations) of stripe patterns [12].

The present letter gives an investigation of the spatio-temporal noise spectra of stripe patterns above the pattern formation threshold. It is shown, by numerical and analytical calculations, that the spatial noise spectra contain sharp peaks (singularities) centered at the wavenumbers of the stripe planform. The temporal power spectra of stripes driven by noise are obtained in a $1/f^a$ form, with $a$ dependent on the spatial dimension of the system. Also the stochastic drift of the stripe patterns is found to be sub-diffusive: whereas usual diffusive drift (e.g. of Brownian particle) obeys a square root law $\sqrt{\langle x(t)^2 \rangle} \propto t^{1/2}$, the stochastic drift of stripe patterns has a root mean wandering $\propto t^{1/4}$ in the case of one spatial dimension.

The numerical analysis in the present letter is performed by solving a stochastic Swift-Hohenberg equation [7]:

$$\frac{\partial A}{\partial t} = pA - A^3 - (\Delta + \nabla^2)^2 A + \Gamma(\mathbf{r},t) \qquad (1)$$

for the temporal evolution of the real-valued order parameter $A(\mathbf{r},t)$, defined in D-dimensional space $\mathbf{r}$. $p$ is the control parameter (the stripe formation instability occurs at $p = 0$), $\Delta$ is the detuning parameter, determining the resonant wavenumber of the stripe pattern $\mathbf{k}_0$: $\mathbf{k}_0^2 = \Delta$, and $\Gamma(\mathbf{r},t)$ is an additive noise, $\mathbf{d}$ - correlated in space and time, and of temperature $T$:
$<\Gamma(\mathbf{r}_1,t_1)\cdot\Gamma(\mathbf{r}_2,t_2)> = 2T\cdot\mathbf{d}(\mathbf{r}_1-\mathbf{r}_2)\mathbf{d}(t_1-t_2)$.

The analytical results are obtained by solving stochastic amplitude equation for stripes:

$$\frac{\partial B}{\partial t} = pB - |B|^2 B - (2i\mathbf{k}_0\nabla + \nabla^2)^2 B + \Gamma(\mathbf{r},t) \qquad (2)$$

for a slowly varying complex-valued envelope $B(\mathbf{r},t)$ of the stripe pattern with the resonant wavevector $\mathbf{k}_0$. The amplitude equation (2) can be obtained directly from (1), by inserting $A(\mathbf{r},t) = (B(\mathbf{r},t)\cdot\exp(i\mathbf{k}_0\mathbf{r}) + c.c.)/\sqrt{3}$, or directly from the microscopic equations of various stripe forming systems (e.g. [13]). (2) can be also written phenomenologically from symmetry considerations for arbitrary stripe patterns [8].

For analytical treatment it is assumed, that the system is sufficiently far above the stripe forming transition: $p \gg T$. Then the homogeneous component $|B_0| = \sqrt{p}$ is dominating in (2) (correspondingly one stripe component in (1) is dominating), and one can look for a solution of (2) in the form of a perturbed homogeneous state: $B(\mathbf{r},t) = B_0 + b(\mathbf{r},t)$. After linearisation of (2) around $B_0$, and diagonalisation, one obtains linear stochastic equations for the amplitude: $b_+ = (b+b^*)/\sqrt{2}$ and phase: $b_- = (b-b^*)/\sqrt{2}$ perturbations:

$$\frac{\partial b_+}{\partial t} = -2pb_+ + \hat{L}_+(\nabla)b_+ + \Gamma_+(\mathbf{r},t) \qquad (3.a)$$

$$\frac{\partial b_-}{\partial t} = \hat{L}_-(\nabla)b_- + \Gamma_-(\mathbf{r},t) \qquad (3.b)$$

where the nonlocality operators $\hat{L}_\pm(\mathbf{k}_0,\nabla)$ are given by:

$$\hat{L}_\pm(\mathbf{k}_0,\nabla) = -p + (2\mathbf{k}_0\nabla)^2 - \nabla^4 \mp \sqrt{p^2 - (4\mathbf{k}_0\nabla^3)^2} \qquad (4.a)$$

and their spectra $L_\pm(\mathbf{k}_0,\mathbf{dk})$ by:

$$L_\pm(\mathbf{k}_0,\mathbf{k}) = -p - (2\mathbf{k}_0\mathbf{dk})^2 - \mathbf{dk}^4 \mp \sqrt{p^2 + (4\mathbf{k}_0\mathbf{dk}^3)^2} \qquad (4.b)$$

as obtained by substitution $\nabla \leftrightarrow i\mathbf{dk}$, where $\mathbf{dk} = \mathbf{k} - \mathbf{k}_0$ is the wavevector of the perturbation mode in (2).

Asymptotically, for $|4\mathbf{k}_0 \nabla^3| \ll p$ the nonlocality operator for phase perturbations $\hat{L}_-(\nabla)$ simplifies to: $\hat{L}_-(\mathbf{k}_0, \nabla) = (2\mathbf{k}_0 \nabla)^2 - \nabla^4$. In the opposite limit, of $|4\mathbf{k}_0 \nabla^3| \gg p$ the nonlocality operator is: $\hat{L}_-(\mathbf{k}_0, \nabla) = -(2i\mathbf{k}_0 \nabla + \nabla^2)^2$.

(3.a) is an equation for amplitude fluctuations $b_+$ corresponding to the modulation amplitude of the stripe pattern. (3.b) is the equation for phase fluctuations $b_-$ corresponding to the parallel translation of the stripes. (3.b) indicates that the phase fluctuations decay with a rate $L_-(\mathbf{k}_0, \mathbf{k}) = -(2\mathbf{k}_0 \mathbf{dk})^2 - \mathbf{dk}^4$ in the strong pump limit, or $L_-(\mathbf{k}_0, \mathbf{k}) = -(2\mathbf{k}_0 \mathbf{dk} + \mathbf{dk}^2)^2$ in the weak pump limit. This means that the long-wavelength phase perturbation modes decay asymptotically slowly, with a decay rate approaching zero for $\mathbf{dk} \to 0$, which is a consequence of the phase invariance of the system.

Next only the phase perturbations are considered. They determine the stochastic dynamics of the stripe pattern above the stripe formation threshold. $p > 0$. More precisely, the amplitude fluctuations are small compared with phase fluctuations if $|4\mathbf{k}_0 \mathbf{dk}^3| \ll p$, as follows from (3).

We calculate spatio-temporal power spectra of phase fluctuations, by rewriting (3.b) in terms of the spatial and temporal Fourier components $b(\mathbf{r}, t) = \int b_-(\mathbf{k}, w) \exp(iwt - i\mathbf{kr}) dw d\mathbf{k}$:

$$S(\mathbf{k}, w) = |b_-(\mathbf{k}, w)|^2 = \frac{|\Gamma_-(\mathbf{k}, w)|^2}{w^2 + |L_-(\mathbf{k}_0, \mathbf{k})|^2} \quad (5)$$

Assuming $\mathbf{d}$ - correlated noise in space and time, $|\Gamma_-(\mathbf{k}, w)|^2$ is simply proportional to the temperature $T$ of the random force.

The spatial power spectrum is obtained by integration of (5) over all spatial frequencies $w$:

$$S(\mathbf{k}) = \int_{-\infty}^{\infty} \frac{T}{w^2 + |L_-(\mathbf{k}_0, \mathbf{k})|^2} dw = \frac{T p}{2|L_-(\mathbf{k}_0, \mathbf{k})|}. \quad (6)$$

This results in a divergence of the spatial spectra at $\mathbf{dk} \to 0$ (equivalently in a divergence of the spatial spectra of roll patterns in (1) for: $\mathbf{k} \to \mathbf{k}_0$). As follows from (6), the perturbations of the stripe pattern $\mathbf{dk}$ diverge differently, depending on whether the perturbations are parallel or perpendicular to the wave-vector of the stripe pattern $\mathbf{k}_0$. This follows from the isotropic form of the nonlocality operator (4). The parallel perturbations (corresponding to compression and undulation of stripes) diverge as $\mathbf{dk}^{-2}$, the perpendicular perturbations (corresponding to zigzaging of stripes) diverge as $\mathbf{dk}^{-4}$. This results in an anisotropic form of the singularity at $\mathbf{dk} = 0$, which can actually be expected from the anisotropic form of the amplitude equation for rolls (2). Fig.1 shows the spatial noise power spectrum of stripe pattern as obtained from numerical integration of SHE (1) and illustrates the anisotropy. The anisotropy results in the stability conditions of stripes for depending on the number of spatial dimensions. Indeed, the integral of (6) over the spatial wavenumbers $\mathbf{dk}$ diverges for spatial dimension $D < 4$, and converges for $D \geq 4$ only. Only for four (and more) dimensions of space are the stripes absolutely stable against additive noise. This is in contrast to a well known theorem concerning the stability of a "condensate": the condensate (a homogeneous distribution) is stable for all spatial dimension larger than two.

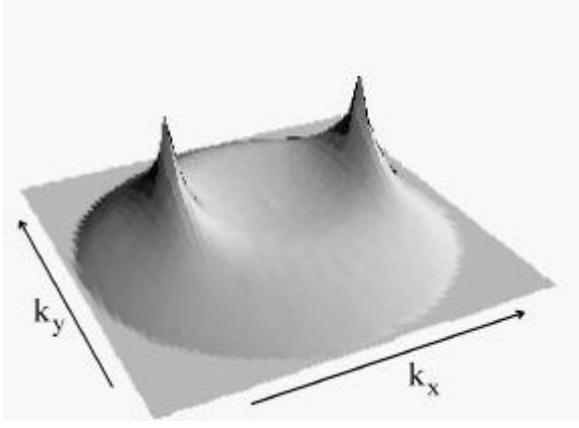

*Fig.1. Spatial noise power spectrum of stripes in 2D obtained numerically by solving stochastic SHE (1) with $p = 1$ and $\Delta = 0.7$. Averaging time $t_{averaging} = 5000$. Intensity of spatial spectral components in logarithmic representation.*

The temporal power spectra are obtained by the integral of (5) over all possible wavevectors $\mathbf{dk}$:

$$S(w) = \int_{-\infty}^{\infty} \frac{T}{w^2 + |L_{-}(\mathbf{k}_0, \mathbf{k})|^2} d\mathbf{k}, \qquad (7)$$

which however has no analytical form, even for one spatial dimension.

Asymptotically, in the limit of small frequencies, $w \to 0$, when the term $(2\mathbf{k}_0 \mathbf{dk})^2$ dominates in the denominator of integral (7), the analytical integration is possible, and leads to the following results. For 1D the spectrum is: $S_{1D}(w) = c_{1D} T/w^{3/2}$, with the coefficient $c_{1D} = p/(2\sqrt{2} k_0^2)$. For 2D: $S_{2D}(w) = c_{2D} T/w^{1.25}$, for 3D: $S_{3D}(w) = c_{3D} T/w$, and in the general case of D dimensions: $S_D(w) = c_D T/w^a$ with $a = 1 + (3-D)/4$ and coefficients $c_D$ of order of unity.

The integral (7) was integrated numerically, and the results for 1,2,3 dimensions are given in Fig.2. $1/w^a$ dependences are obtained. In the small frequency limit $w \to 0$, the exponents obey $a = 1 + (3-D)/4$; in the large frequency limit $w \to \infty$ the spectra show also a power law form, however with exponents $a = 1 + (4-D)/4$. The exponents change abruptly from small frequency values $a = 1 + (3-D)/4$ to large frequency values $a = 1 + (4-D)/4$ at the critical frequency $w_{cr} \approx 4\mathbf{k}_0^2$, as follows from the analysis of (4), and as seen from Fig.2

In this way 1/f spectra are obtained for fluctuations of stripes subjected to additive white noise. We note, that the 1/f-, or "flicker" noise is an old puzzle of physics: It is found in many different kinds of systems, from physics, technology, biology, astrophysics, geophysics and sociology [14]. Despite its ubiquity, a universal explanation of flicker noise has never been given. Recently such 1/f noise was obtained for "condensates", described by a stochastic Ginzburg-Landau equation [15], where the exponent $a$ depends on the dimension of space as $a = 1 + (2 - D)/2$. Comparing with the results of [15] one can conclude, that:

1) One-dimensional stripes have the same exponent of noise power spectra as one dimensional condensates. This is plausible, since the amplitude equation for stripes is similar to a complex Ginzburg-Landau equation, and the two coincide in the limit of $\mathbf{dk} \to 0$.

2) Two-dimensional stripes behave like noisy condensates of dimension of $D = 1.5$, if one judges from the exponents of the noise spectra in the low frequency limit. As discussed above (see also Fig.1) the singularity of spatial noise spectra is strongly squeezed in the direction along the stripes. It is then plausible, that the noise characteristics of this anisotropic system are obtained in between those for a one- and isotropic two-dimensional system.

3) Similarly three-dimensional stripes (lamellae) behave like two-dimensional condensates. Both display power spectra with $a = 1$.

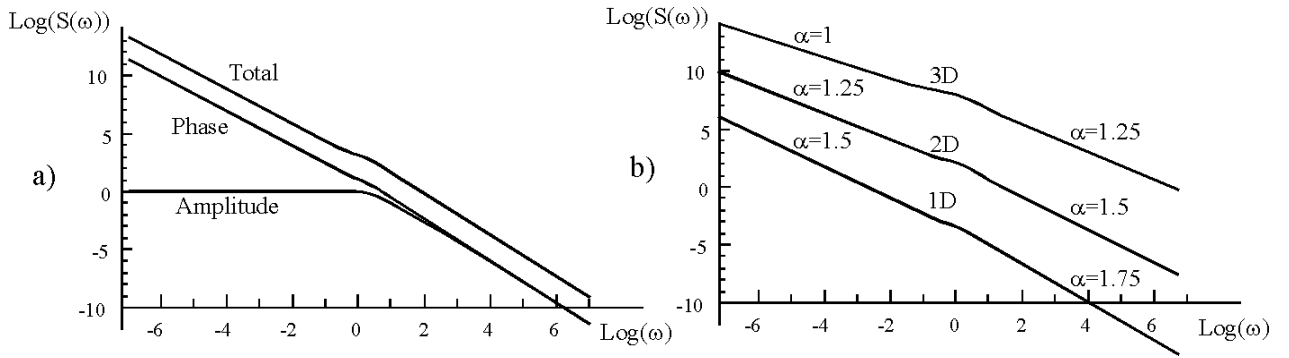

*Fig.2.* Temporal spectra as obtained by numerical calculation of the integral (7) with $p = 1$, and $\mathbf{k}_0^2 = \Delta = 1$: a) 1D case. The phase power spectrum (as following from integration of (3.b)), the amplitude power spectrum (as following from integration of (3.a)), and the total spectrum is shown. b) The phase power spectra as calculated for 1,2, and 3 spatial dimensions.

The integral of the $1/f^a$ power spectrum always diverges in the limit of large or of small frequency, indicating a breakup of the ordered state in the limit of small or of large times, respectively. For the stripes in up to three spatial dimensions $a \geq 1$ the integral of the temporal power spectra diverges at low frequencies, which means that the average size of the fluctuations of the order parameter grows to infinity for large times. The average size of fluctuations is: $\left\langle |b(t)|^2 \right\rangle \approx \int_{w_{\min}}^{\infty} S(w) dw$, where $w_{\min} \approx 2p/t$ is the lower cut-off boundary of the temporal spectra. Thus the variance of the order parameter for processes with $1/f^a$ noise spectra grows as $\left\langle |b(t)|^2 \right\rangle \propto t^{a-1}$ with increasing time. This generalizes the well known Wiener stochastic diffusion process obeying a linear diffusion law for the variance (or equivalently a square root law $\sqrt{\left\langle x(t)^2 \right\rangle} \propto t^{1/2}$ for the root mean wandering). The Wiener law is well known for zero dimensional systems, e.g. Brownian motion. From our results it follows that diffusion processes in spatially extended systems are weaker than in zero dimensional systems. In particular the variance of the order parameter in 1D systems ($a = 1.5$) increase as $\langle |b(t)|^2 \rangle \propto t^{1/2}$. We tested this stochastic drift of stripe pattern by numerically solving the Swift-Hohenberg equation (1) in 1D. We calculated the displacement of the stripe pattern as a function of time. (The displacement $x(t)$ of the stripe position in SHE (1) is directly proportional to the phase of the order parameter $B(x,t)$ at the corresponding spatial location in amplitude equation (2).) Fig.3.a shows the power spectrum of displacement, which follows a $w^{-3/2}$ law, in accordance with the analytical predictions. Fig.3.b shows the power spectrum of the variation (temporal derivative $x(t) - x(t - \Delta t)$) of the displacement, which follows a $w^{1/2}$ law respectively. The average square displacement of stripe position $x(t)$, as averaged over many realizations is shown in Fig.3.c. The predicted slope of 1/2 is clearly seen for times up to $t \approx 1000$. For very large times the usual (Brownian) stochastic drift is obtained. This behaviour for large times (correspondingly small frequencies) however is an artifact of numerical space discretization. A sub-diffusive stochastic drift of kinks (fronts) in 1D Ginzburg-Landau equation (for small times, however) was recently found in [16].

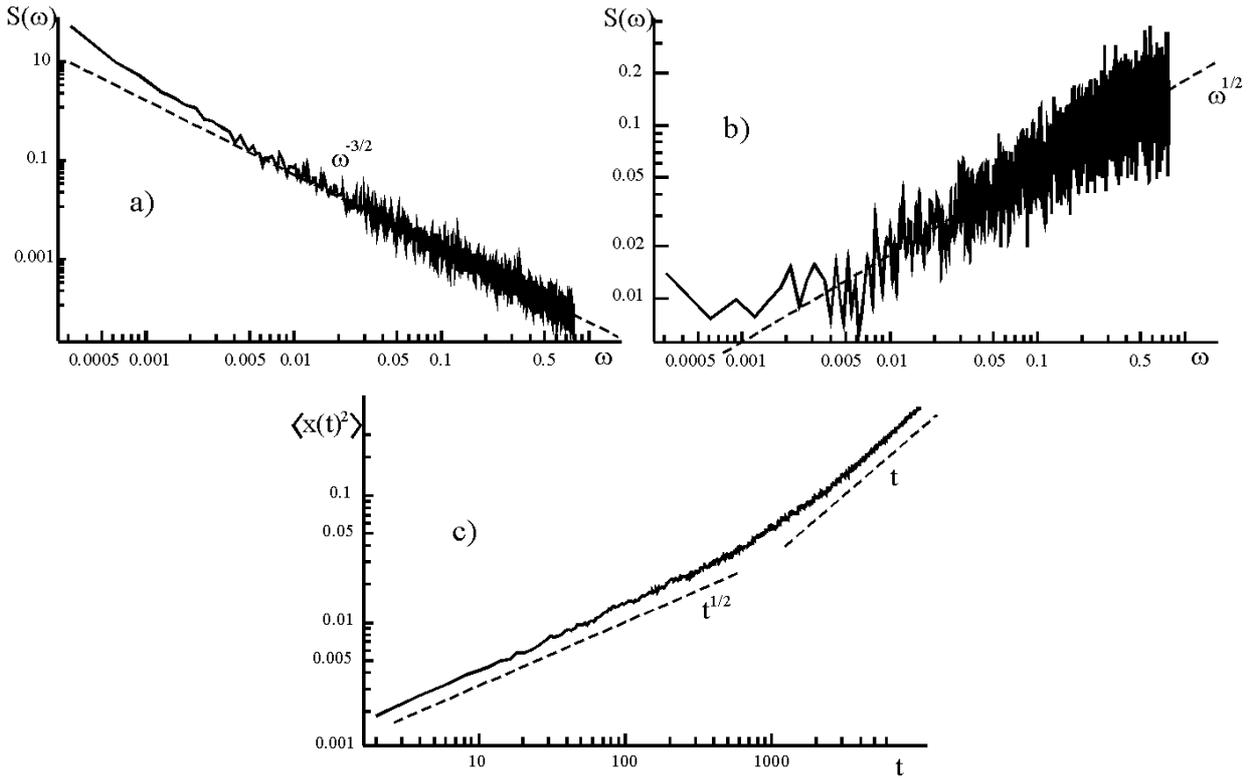

*Fig.3. Statistical properties of position of stripe pattern as obtained by numerical integration of SHE (1) in 1D with $p = 1$ and $\Delta = 0.7$. a) the power spectrum of the displacement $x(t)$. The dashed line with slope $w^{-3/2}$ serves to guide the eye. b) the power spectrum of variation of the displacement $x(t) - x(t - \Delta t)$. Dashed line with slope $w^{1/2}$ serves to guide the eye. c) the average square displacement of stripe position $x(t)$ as averaged over 1000 realizations.*

The above discussion on stochastic drifts concerns large times: the variance of the position of 1D stripes $t^{1/2}$ is related with the $w^{-1.5}$ power spectrum at small frequencies. The $w^{-1.75}$ spectrum at large frequencies ($w \geq w_{cr} \approx 4\mathbf{k}_0^2$) predicts equally a $t^{3/4}$ law for the stochastic drift at small times. Our numerical calculations in Fig.3 do not consider, however, the small time scales ($t \leq 2p/w_{cr}$), thus the small time drift law was not numerically observed.

Our analysis in 2D predicts the stochastic drift obeying a $t^{1/4}$ law for large times, and a $t^{1/2}$ for small times.

The stochastic drift (although sub-diffusive) of the order parameter means that for large times the fluctuations become on average of the order of magnitude of the order parameter itself. The long range order eventually breaks up even for a small temperature. In general for $1/f^a$ power spectra with $a > 1$ such finite perturbations occur for times $t \geq t_{cr} \propto T^{-1/(a-1)}$. We tested this dependence on 1D stripes, where the critical time is $t_{cr} \propto T^{-2}$. For this purpose we prepared numerically an off-resonance stripe pattern in SHE (1) for 1D without stochastic term. The off-resonance stripe was stable (was within the Eckhaus stability range). Then we switched on the stochastic term and waited until the fluctuations of the stripe pattern grew and destroyed the stripe pattern locally. After the stripe pattern is destroyed in some place, a resonant stripe pattern appears there and invades the whole pattern in the form of propagating switching waves. The state of the system in this way changes from a local potential minimum (off-resonance stripe) to the global potential minimum (resonant stripe) as triggered by a local perturbation.

In Fig.4 the numerically calculated life times of the off-resonance stripe pattern are plotted, depending on the temperature of the stochastic force. Again, as predicted by analytic calculations the dependence $t_{cr} \propto T^{-2}$ is obtained. This shows that in spatially extended systems the switching time from the local potential minimum to the deeper global minimum does not depends exponentially on time as in zero-dimensional (compact) systems, but obeys a power law. In particular for stripe patterns the switching time is: $t_{switch} \propto T^{-2}$ in 1D, and $t_{switch} \propto T^{-4}$ in 2D.

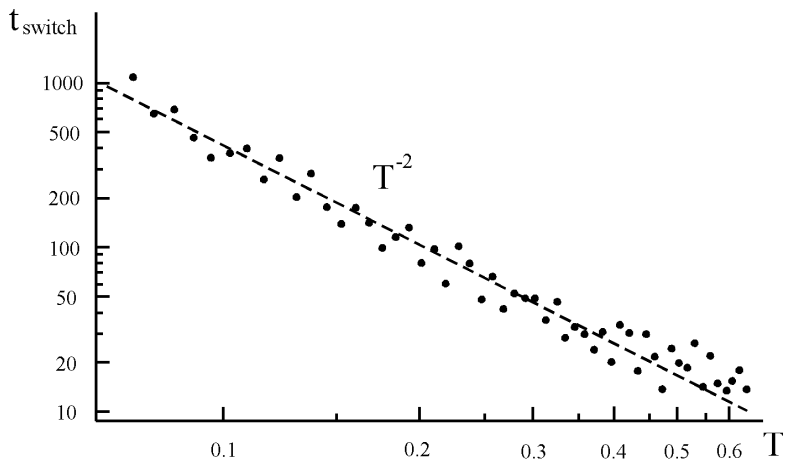

*Fig.4. Life-time of off-resonant stripe pattern, depending on the noise temperature T, as obtained by numerical integration of SHE (1) in 1D with $p = 1$. The resonant stripe pattern with $\mathbf{k}_0^2 = 1$ was excited for $\Delta = 1$. The detuning value was then reduced to 0.75, and the time was measured until the new resonant stripe pattern wins. Every point is obtained by averaging over 10 realizations. The dashed line with slope $T^{-2}$ serves to guide the eye.*

Concluding: simple models for stripe patterns (stochastic Swift-Hohenberg equation for order parameter, and stochastic Newell-Whitehead-Segel equation for the envelope of stripes) allow to calculate spatio-temporal noise power spectra, and to predict the following properties of the stripe patterns in presence of noise:

1) anisotropic form of singularities in the spatial power spectra;

2) stability conditions dependent on number of spatial dimensions;

3) $1/f^a$ temporal power spectra with the exponent $a$ depending explicitly on the number of spatial dimensions;

4) sub-brownian of stochastic drift law;

5) power law temperature dependence of life-time of locally stable stripe patterns (corresponding to local potential minimum).

Discussions with C.O.Weiss are acknowledged. This work has been supported by Sonderforschungs Bereich 407, and by ESF Network PHASE.